\documentclass[default,iicol,sn-mathphys]{sn-jnl}
\usepackage{color}

\UseRawInputEncoding
\begin{document}
 
\title[ ]{Observation of superconducting diode effect in antiferromagnetic Mott insulator $\alpha$-RuCl$_3$}


\author[1,2]{Jiadian He}
\equalcont{These authors contributed equally to this work.}
\author[1,2]{Yifan Ding}
\equalcont{These authors contributed equally to this work.}
\author[1,2]{Xiaohui Zeng}
\author[1,2]{Yiwen Zhang}
\author[1,2]{Yanjiang Wang}
\author[1,2]{Peng Dong}
\author[1,2]{Xiang Zhou}
\author[1,2]{Yueshen Wu}
\author[1]{Kecheng Cao}
\author[3]{Kejing Ran}
\author*[1,2]{Jinghui Wang}\email{wangjh2@shanghaitech.edu.cn}
\author[1,2,4]{Yulin Chen}
\author[5]{Kenji Watanabe}
\author[6]{Takashi Taniguchi}
\author[7,8]{Shun-Li Yu} 
\author[7,8]{Jian-Xin Li} 
\author*[7,8]{Jinsheng Wen} \email{jwen@nju.edu.cn}
\author*[1,2]{Jun Li} \email{lijun3@shanghaitech.edu.cn}

\affil[1]{\orgdiv{School of Physical Science and Technology}, \orgname{ShanghaiTech University}, \orgaddress{\city{Shanghai}, \postcode{201210}, \country{China}}}

\affil[2]{\orgdiv{ShanghaiTech Laboratory for Topological Physics}, \orgname{ShanghaiTech University}, \orgaddress{\city{Shanghai}, \postcode{201210}, \country{China}}}

\affil[3]{\orgdiv{College of Physics \& Center of Quantum Materials and Devices}, \orgname{Chongqing University}, \orgaddress{\city{Chongqing}, \postcode{401331}, \country{China}}}

\affil[4]{\orgdiv{Department of Physics, Clarendon Laboratory}, \orgname{University of Oxford}, \orgaddress{\city{Oxford}, \postcode{OX1 3PU}, \country{UK}}}

\affil[5]{\orgdiv{Research Center for Functional Materials}, \orgname{National Institute for Materials Science}, \orgaddress{\city{Tsukuba}, \postcode{305-0044}, \country{Japan}}}

\affil[6]{\orgdiv{International Center for Materials Nanoarchitectonics}, \orgname{National Institute for Materials Science}, \orgaddress{\city{Tsukuba}, \postcode{305-0044}, \country{Japan}}}

\affil[7]{\orgdiv{School of Physics}, \orgname{Nanjing University}, \orgaddress{\city{Nanjing}, \postcode{210093}, \country{China}}}

\affil[8]{\orgname{Collaborative Innovation Center of Advanced Microstructures}, \orgaddress{\city{Nanjing}, \postcode{210093}, \country{China}}}

\abstract{
Nonreciprocal superconductivity, also called as superconducting diode effect that spontaneously breaks time-reversal symmetry, is characterized by asymmetric critical currents under opposite applied current directions. This distinct state unveils a rich ore of intriguing physical properties, particularly in the realm of nanoscience application of superconductors. Towards the experimental realization of superconducting diode effect, the construction of two-dimensional heterostructures of magnets and $s$-wave superconductors is considered to be a promising pathway. In this study, we present our findings of superconducting diode effect manifested in the magnetic Mott insulator $\alpha$-RuCl$_3$. This phenomenon is induced by the proximity effect within a van der Waals heterostructure, consisting of thin $\alpha$-RuCl$_3$/NbSe$_2$ flakes. Through transport property measurements, we have confirmed a weak superconducting gap of 0.2~meV, which is significantly lower than the intrinsic gap of NbSe$_2$(1.2~meV). Upon the application of a weak magnetic field below 70~mT, we observed an asymmetry in the critical currents under positive and negative applied currents. This observation demonstrates a typical superconducting diode effect in the superconducting $\alpha$-RuCl$_3$. The superconducting diode effect and nonreciprocal resistance are observed exclusively when the magnetic field is aligned out-of-plane. This suggests that an Ising-type spin-orbit coupling in the superconducting $\alpha$-RuCl$_3$ may be responsible for the mechanism. Our findings furnish a platform for the exploration of superconducting diode effect via the artificial construction of heterostructures.
}

\maketitle

Nonreciprocal superconductivity, characterized by a non-dissipative superconducting current, permits the flow of supercurrent in one direction, while allowing a normal current to flow in the opposite direction. This phenomenon indicates a diode effect-like behavior, and is thus generally called as superconducting diode effect (SDE)\cite{ando2020observation,narita2022field,nadeem2023superconducting,jiang2022superconducting,mao2024universal,yuan2022supercurrent,wu2022field,Zhang2023,bauriedl2022supercurrent,lin2022zero,ghosh2024high,lyu2021superconducting,le2024superconducting,PhysRevB.108.064501}. The remarkable characteristic of SDE offers substantial potential for the application of chiral-like superconductors in the fields of superconducting electronics, superconducting spintronics, and quantum information and communication technology. In SDE, both inversion and time-reversal symmetry should be simultaneously broken, thus, it is often observed in the two-dimensional superconducting systems with strongly spin-orbital coupling (SOC). The first discoveryjwen of SDE is on a superconducting [Nb/V/Ta]$_n$ superlattice \cite{ando2020observation}, and its subsequent research achieving a field-free SDE on the [Nb/V/Co/V/Ta] superconducting films by importing ferromagnet multilayers \cite{narita2022field}. Another notable progess is based on Josephson junctions due to the asymmetric Josephson tunneling, including NbSe$_2$/Nb$_3$Br$_8$/NbSe$_2$ hybrid Josephson junction \cite{wu2022field} and the twisted Bi$_2$Sr$_2$CaCu$_2$O$_{8+\delta}$ (BSCCO) \cite{ghosh2024high}. The twisted graphene also demonstrates a SDE owing to the inversion symmetry breaking \cite{lin2022zero}. Moreover, SDE has been discovered in some unconventional superconductivity under spontaneously time-reversal symmetry breaking, including the NbSe$_2$ with Ising superconductivity nature \cite{bauriedl2022supercurrent,Zhang2023}, and topological superconductor of CsV$_3$Sb$_5$ \cite{le2024superconducting,wu2022nonreciprocal}. 

Recently, an antiferromagnetic (AFM) Mott insulator $\alpha$-RuCl$_3$ has been intensively studied for its exotic quantum properties \cite{PhysRevLett.114.147201,ran2017prl} and potential for inducing nonreciprocal superconductivity. In $\alpha$-RuCl$_3$, each Ru ion has a hole, which occupies a spin-orbit-coupled $j_{\text{eff}}$ = 1/2 ground state. This state is strongly localized due to Coulomb interactions, resulting in a spin-orbit-assisted Mott insulating phase and a subsequent zigzag AFM order. Therefore, it is highly plausible to observe exotic superconducting states by inducing charge doping into $\alpha$-RuCl$_3$. In the honeycomb lattice, the spin interaction could be described by a minimal $K-\Gamma$ effective-spin model that includes the direction-dependent Kitaev coupling $K$ term and isotropic AFM off-diagonal exchange interaction $\Gamma$ term \cite{PhysRevB.96.115103,ran2017prl}. This suggests that charge doping, in conjunction with magnetic coupling, may induce unconventional superconductivity that spontaneously breaks time-reversal symmetry \cite{bbwang2019cpb}. Nevertheless, efforts of chemically doped $\alpha$-RuCl$_3$ have not yet been successful. This is primarily due to the considerable challenge posed by the large electronegativity of Cl$^-$ ions, which could significantly trap or exclude carriers. Although there have been reports of heavily doped potassium (K$_{0.5}$RuCl$_3$) \cite{PhysRevMaterials.1.052001}, the samples remain insulating and appear to possess a charge-ordered ground state. Furthermore, the application of high pressure \cite{PhysRevB.97.245149} or electric field gating \cite{PhysRevB.100.165426,mashhadi2019spin} has proven to be ineffective in modifying the band structure into a conducting state. Alternatively, a heterostructure between a few layers of $\alpha$-RuCl$_3$ and a superconductor substrate may present a promising approach to induce unconventional superconductivity in $\alpha$-RuCl$_3$ via proximity effects from $s$-wave superconductors like few-layer NbSe$_2$  \cite{narita2022field,muller2021temperature,kezilebieke2020topological,jo2023local,amundsen2024colloquium}.  Given that a monolayer or few-layer $\alpha$-RuCl$_3$ naturally breaks inversion symmetry, it could, in conjunction with spontaneously broken time-reversal symmetry, induce various novel physical behaviors. Especially, the superconductivity effect that induces effective pairing near the single Fermi pocket can be considered equivalent to the performance of SDE \cite{mao2024universal,yuan2022supercurrent,yi2022crossover,wan2023orbital,ding2023constructing}.

In the present study, we investigated the emergence of nonreciprocal superconductivity in the Mott insulator $\alpha$-RuCl$_3$, which could be induced by the superconducting proximity effect from NbSe$_2$. The device was engineered as a van der Waals heterostructure with a layered $\alpha$-RuCl$_3$/NbSe$_2$ configuration, as depicted in the optical image in Fig.\ref{fig1}(a). The device was specifically engineered to enable the measurement of transport properties of $\alpha$-RuCl$_3$ crystals ranging from monolayer to few-layer thickness. From differential resistance measurements, an emergent superconducting gap of 0.17 meV at 1.8 K was observed in $\alpha$-RuCl$_3$. This gap is significantly smaller than the intrinsic superconducting gap of NbSe$_2$, which is 1.2 meV. Upon the application of a weak magnetic field less than 70~mT, an asymmetry in the critical currents was observed under positive and negative excitation currents. This phenomenon signifies the typical SDE in the superconducting $\alpha$-RuCl$_3$.

\subsection*{Results and discussion}
\begin{figure*}[!htbp]
\includegraphics[width=0.8\textwidth,clip]{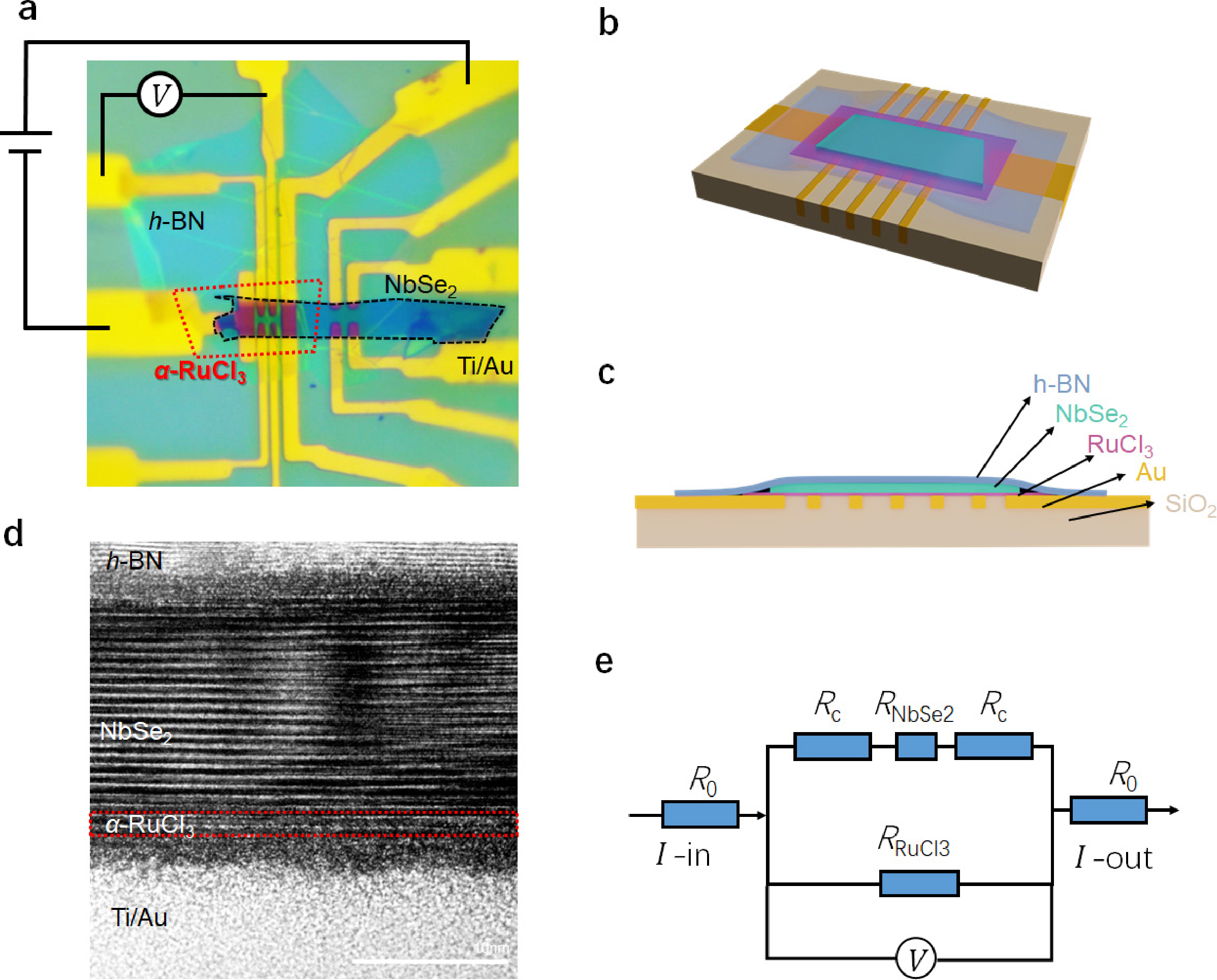}
\centering
\caption{\label{fig1} \textbf{Schematic structure and microscopic characterization of $\alpha$-RuCl$_3$/NbSe$_2$ device.} \textbf{(a)} Optical image of $\alpha$-RuCl$_3$/NbSe$_2$ device. The stacked$\alpha$-RuCl$_3$, NbSe$_2$ and $h$-BN are delineated by red dotted lines, black dashed lines, and white dashed lines, respectively. \textbf{(b)} Schematic overview of $\alpha$-RuCl$_3$/NbSe$_2$ device. \textbf{(c)}  Cross-sectional view of the heterostructure device. \textbf{(d)} High-resolution STEM cross section image of the device showing the well-defined interfaces of the heterostructure, the scale bar is 10 nm. \textbf{(e)} The abstracted circuit of the device. The measured current consists of two contributions from the parallel circuits in few-layer $\alpha$-RuCl$_3$ and NbSe$_2$ flake with contact resistance $R_c$.
}
\end{figure*}

The schematic structure of $\alpha$-RuCl$_3$/NbSe$_2$ device is illustrated in Fig.\ref{fig1}(b) and (c).
Ti/Au electrodes were embedded onto SiO$_2$/Si substrates to establish a gentle yet firm contact. In contrast, electrodes fabricated using the standard lithography process often protrude significantly, which would damage the delicate few-layer $\alpha$-RuCl$_3$ or NbSe$_2$ due to abrupt height transitions at the edges, potentially leading to a current shortage through NbSe$_2$. The bottom layers of $\alpha$-RuCl$_3$ are selected and positioned to fully cover the electrodes and isolate them from NbSe$_2$. An overlying $h$-BN flake is applied to shield the device from degradation.

As a heterostructure device, the quality of the interfaces between Au/Ti electrodes, $\alpha$-RuCl$_3$, NbSe$_2$, and even $h$-BN is crucial to the transport properties. Scanning transmission electron microscopy (STEM) has been utilized to investigate the interfaces, as depicted in the cross-sectional view shown in Fig.\ref{fig1}(d). The STEM image reveals high-quality interfaces between Au-$\alpha$-RuCl$_3$, $\alpha$-RuCl$_3$ -NbSe$_2$, and also NbSe$_2$-BN, all of which display van der Waals contact with no observable bubbles or degradation in the images. For this specifically structured device, the abstracted circuit is illustrated as shown in Fig.\ref{fig1}(e). In this configuration, the current flows from both $\alpha$-RuCl$_3$ and NbSe$_2$ in a parallel connection. Given that the interface between the insulating $\alpha$-RuCl$_3$ and the metallic NbSe$_2$ should behave as a Schottky junction, the contact resistance is significantly smaller, because the tunnelling junction resistance between few layer RuCl$_3$ and NbSe$_2$ is only tens of ohm as demonstrated in the Supplementary Imformation. Considering their different work functions\cite{Work_Function_Engineering,Shimada_1994}, there could be a sizable charge transfer between $\alpha$-RuCl$_3$ and NbSe$_2$, which dopes $\alpha$-RuCl$_3$ and turns it into conducting states.
Nevertheless, the conducting state of a monolayer or fewer than 7 layers of $\alpha$-RuCl$_3$ could be drastically altered by the strong coupling of the superconducting wave function from NbSe$_2$, a phenomenon known as the superconductivity proximity effect. 

\begin{figure*}[!htbp]
\includegraphics[width=0.8\textwidth,clip]{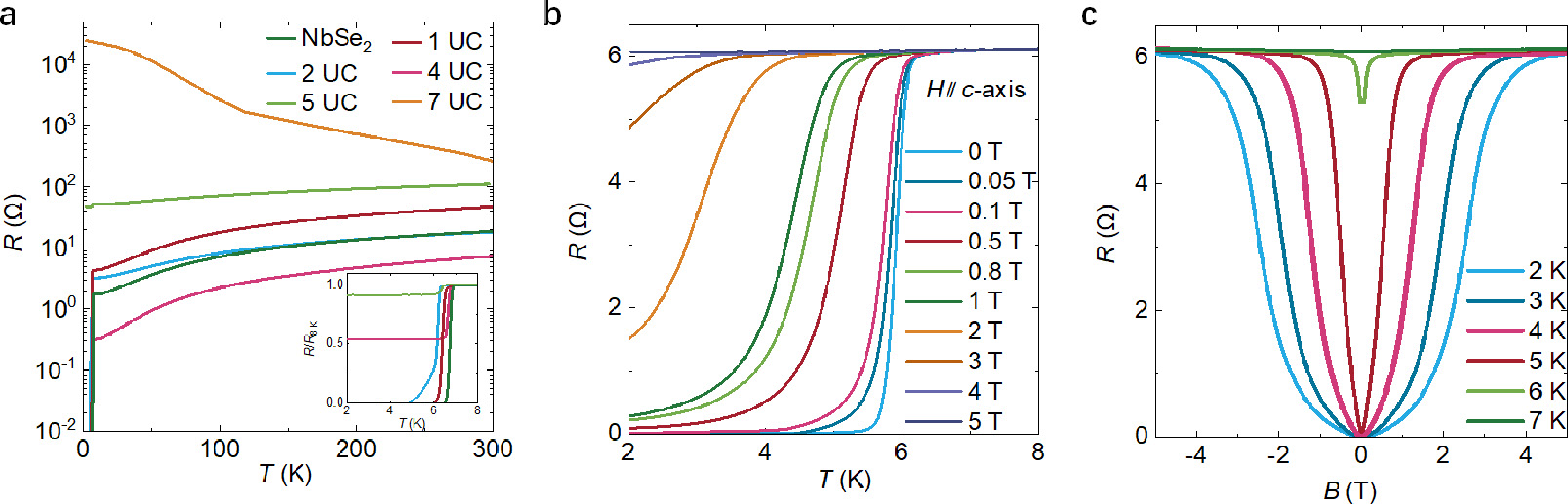}
\centering
\caption{\label{fig2} \textbf{Superconducting characterization of the $\alpha$-RuCl$_3$/NbSe$_2$ heterostructure.} \textbf{(a)} Temperature dependence of the $\alpha$-RuCl$_3$/NbSe$_2$ heterostructure, where $\alpha$-RuCl$_3$ varies in different thickness. \textbf{(b)} Resistance of monolayer $\alpha$-RuCl$_3$/NbSe$_2$ as a function of temperature under out-of-plane magnetic fields from 0 T to 4 T. \textbf{(c)} Magnetoresistance at different temperatures for monolayer $\alpha$-RuCl$_3$/NbSe$_2$ heterostructure.
}
\end{figure*}

Introducing superconductivity into magnetic materials via the superconductivity proximity effect presents a significant challenge, particularly for ferromagnets. This is due to the markedly low diffusive limit of the penetration depth ($\xi_M$) of Cooper pairs into the magnet, given by $\xi_M$ = $\sqrt{D_M/2h} $, where $D_{M}$ is the diffusion coefficient of the magnet, and $h$ is the exchange field in the magnet. Consequently, the $\xi_M$ is typically as low as about 1 nm or even less for a strongly coupled magnetism, and the superconducting wave function decays exponentially, with $\Delta \sim$ exp($D_M/\xi_M$). To explore the diffusive penetration depth of AFM $\alpha$-RuCl$_3$, we examined the transport properties of the $\alpha$-RuCl$_3$/NbSe$_2$ heterostructure with varying thickness of $\alpha$-RuCl$_3$. Fig.\ref{fig2}(a) presents the temperature-dependent resistance of  $\alpha$-RuCl$_3$ at different thicknesses. Firstly, we observe two distinct superconducting transitions for each $R-T$ curve. The initial transition ($T_c^1$) around 6~K aligns with the superconducting transition of pure NbSe$_2$. Conversely, the subsequent transition ($T_c^2$) occurring at a relatively lower temperature corresponds to another superconducting state. Importantly, this transition is dependent on the thickness of $\alpha$-RuCl$_3$. When the $\alpha$-RuCl$_3$ layer ($\sim$ 4 layers) exceeds the superconducting coherence length, the charge transfer between $\alpha$-RuCl$_3$ and NbSe$_2$ induces electrical transport in few layers of $\alpha$-RuCl$_3$. This results in a residual resistance following the superconducting transition. Notably, as the thickness of $\alpha$-RuCl$_3$ continues to increase, the $R-T$ curve exhibits an obvious metal-insulator transition at low temperatures, and the signal indicative of the superconducting transition disappears entirely.

In our experiments, the thickness of the NbSe$_2$ flake, as characterized by Transmission Electron Microscopy (TEM) and Atomic Force Microscopy (AFM), is approximately 15$\sim$20 nm. The temperature dependence of the NbSe$_2$ resistance behaves similarly to that of a bulk crystal, as extensively studied in previous work\cite{xi2016ising}. Hence, the superconducting transition in the $\alpha$-RuCl$_3$ layers should be associated with the NbSe$_2$ flake, considering the superconducting proximity effects. Simultaneously, we can confirm that the superconducting signal originates from $\alpha$-RuCl$_3$, rather than being short-circuited by NbSe$_2$, a topic that is discussed in detail in the Supplementary Information.

Additionally, we investigated the magnetic field dependence of superconductivity for the monolayer $\alpha$-RuCl$_3$/NbSe$_2$ device, as depicted in Fig.\ref{fig2}(b) and Fig.\ref{fig2}(c). Compared with the pure NbSe$_2$, monolayer $\alpha$-RuCl$_3$/NbSe$_2$ exhibits a more sensitive response to the out-of-plane magnetic field. With a modest magnetic field of 0.05~T, the $T_c^2$ of the $R-T$ curve exhibits a noticeable decrease until it completely disappears at a magnetic field of 0.5~T, indicative of the proximity-induced superconductivity. Superconductivity is entirely suppressed under a 5~T magnetic field, which is derived from the intrinsic superconductivity of NbSe$_2$. Consistently, the magnetoresistance of monolayer $\alpha$-RuCl$_3$/NbSe$_2$ at varying temperatures suggests a decrease in the critical field with increasing temperature. Due to the presence of weak proximity-induced superconductivity, a finite magnetoresistance arises in a minor field and escalates with the magnetic field. As a result, the magnetoresistance displays an approximate V-shaped pattern, even at lower temperatures.

\begin{figure*}[!htbp]
\includegraphics[width=0.8\textwidth,clip]{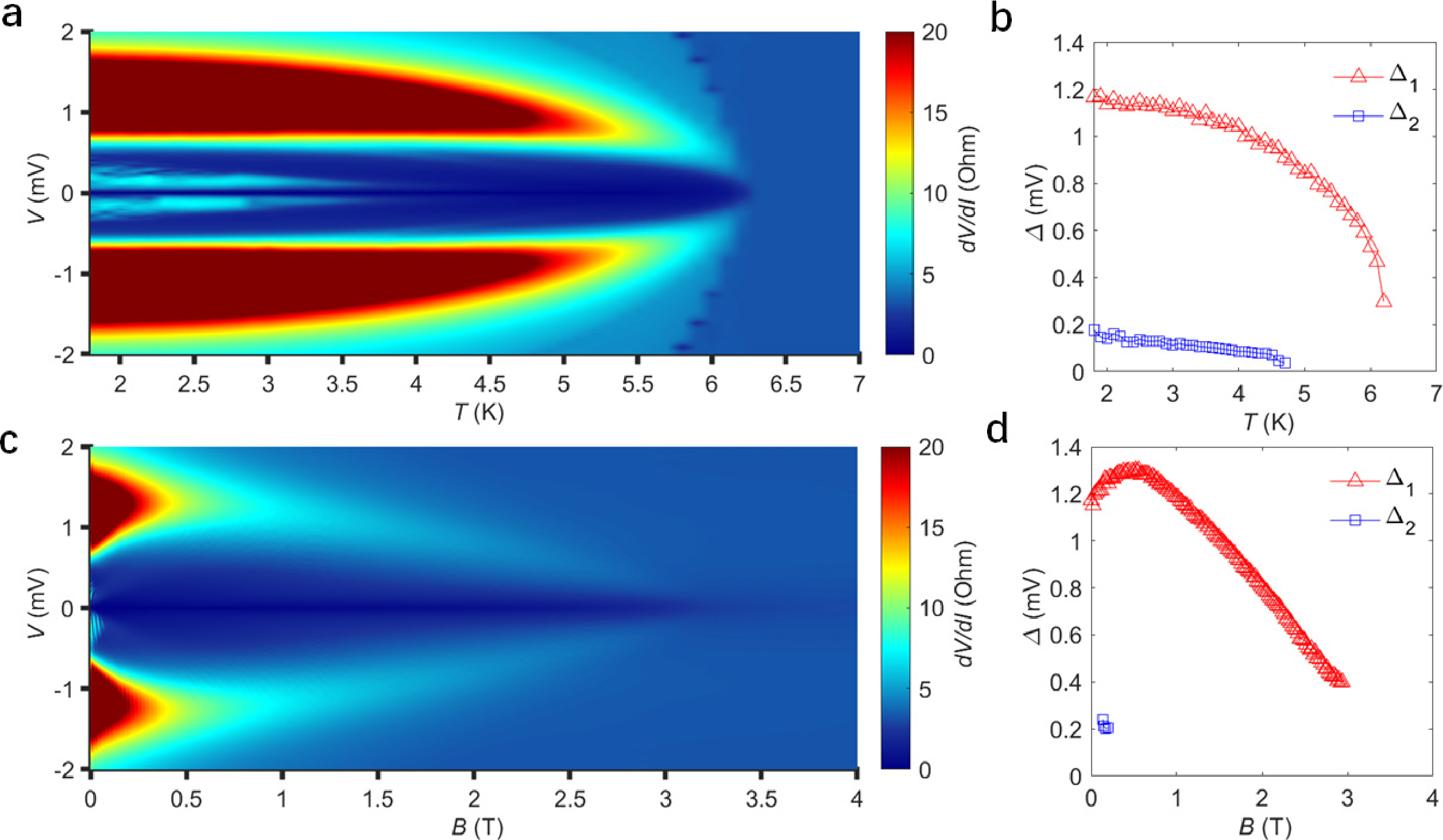}
\centering
\caption{\label{fig3} \textbf{Evolution of differential resistance (d$V$/d$I$) with temperature and magnetic field.} \textbf{(a)} Contour mapped $dV/dI$ spectra with temperature under zero field. A couple of peaks emerge below the superconducting transition temperature of NbSe$_2$. \textbf{(b)} Dependence of superconducting gaps of the pure NbSe$_2$ ($\Delta_0$) and $\alpha$-RuCl$_3$ ($\Delta_1$) with temperature, extracted as the peak positions of the corresponding $dV/dI$ curves. \textbf{(c)} Contour map of $dV/dI$ spectra with magnetic field at 2~K, two pair of peaks could be observed.  \textbf{(d)} Dependence of superconducting gaps with magnetic fields, extracted as the peak positions from each d$V$/d$I$ curve.
}
\end{figure*}

To delve further into the induced superconductivity in $\alpha$-RuCl$_3$, we examined the differential resistance (d$V$/d$I$), measured across a range of temperatures and magnetic fields, as depicted in Fig. \ref{fig3}. In the d$V$/d$I$ spectrum at 2~K, a pair of prominent peaks are observed at $V_0$ = 1.2 mV, which can be estimated as the superconducting gap of NbSe$_2$ ($\Delta_0\sim$ 1.2~meV). The gap progressively disappears above 6.2~K as illustrated in Fig. \ref{fig3} (b), behaving as a typical $s$-wave superconductivity describing by the Bardeen-Cooper-Schrieffer (BCS) model. The temperature-dependent superconducting gaps ($\Delta_1$) are an order of magnitude smaller than those of NbSe$_2$ (see Fig.\ref{fig3} (b)) and disappear at around 4.5 K, which is also lower than that of NbSe$_2$. This should correspond to the proximity-induced superconductivity gap as previously discussed.

In the differential resistance spectra under magnetic fields at 2 K, as depicted in Fig.\ref{fig3} (c) and (d), the $\Delta_0$ is suppressed by fields exceeding 3 T, which aligns with the intrinsic behavior of NbSe$_2$. However, the proximity-induced gap $\Delta_1$ is significantly sensitive to magnetic fields. A field as low as 0.14~T can completely suppress superconductivity (see Fig.\ref{fig3} (d)), indicating an extremely low upper critical field $H_{c2}$. Interestingly, below $H_{c2}$, the differential resistance spectrum displays an asymmetric profile, potentially signifying the presence of chiral superconductivity in $\alpha$-RuCl$_3$.

Based on these findings, we can confidently infer the presence of proximity effect-induced superconductivity within monolayer and few-layer $\alpha$-RuCl$_3$. (A direct substantiation of the induced superconductivity comes from the tunneling junction, as detailed in Fig. SXX.) However, once the current from the tunneling junction becomes the dominant factor in the measurement, it can be observed that the superconductivity should be entirely attributed to the pure NbSe$_2$. This is because the supercurrent will flow across the junction and NbSe$_2$, bypassing the $\alpha$-RuCl$_3$. Consequently, it becomes challenging to explain the existence of a considerably low subgap.

\begin{figure*}[!htbp]
\includegraphics[width=0.8\textwidth,clip]{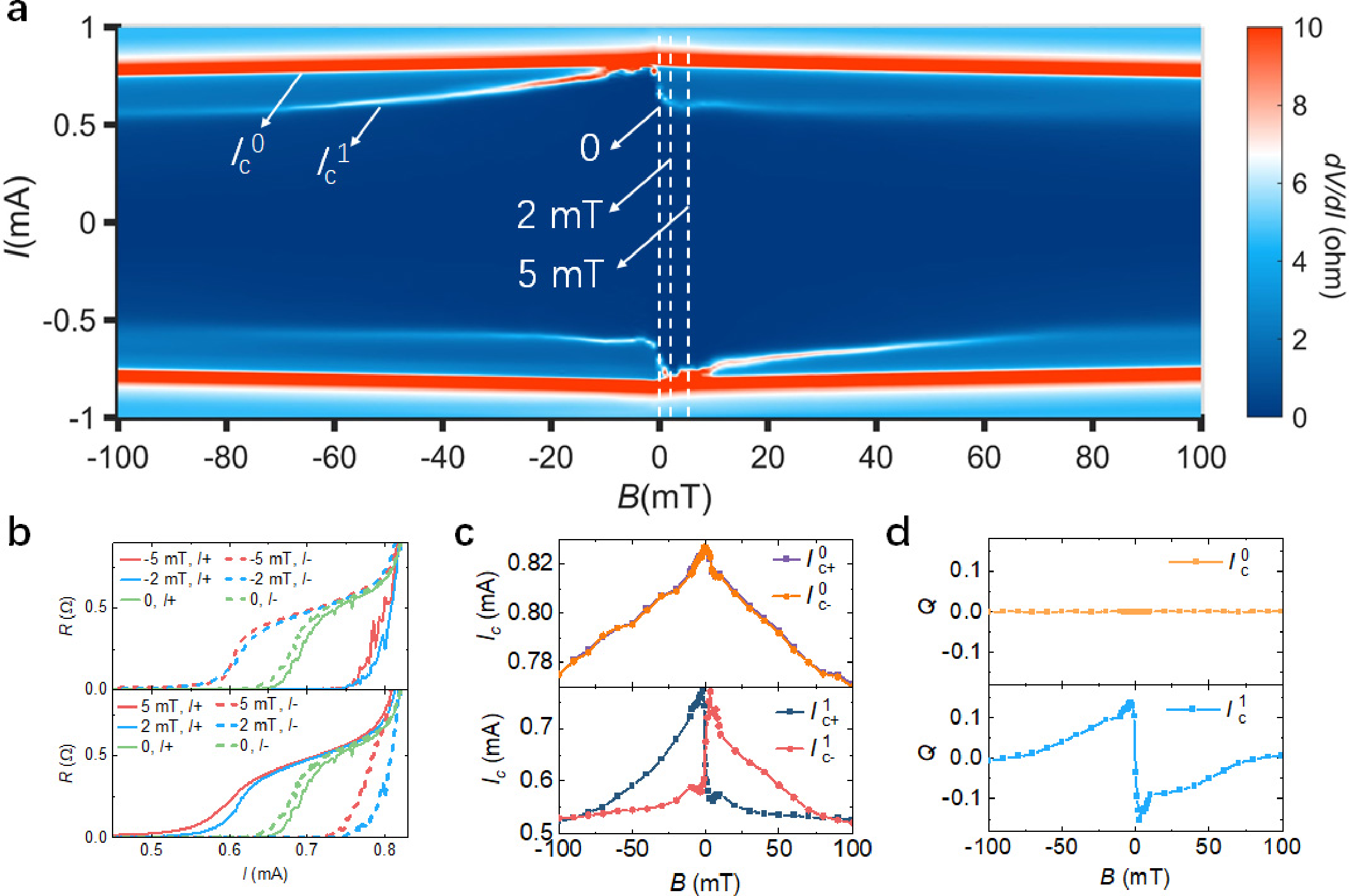}
\centering
\caption{\label{fig4} \textbf{Characterization of chiral superconducting state} \textbf{(a)} Mapping of d$V$/d$I$ under magnetic fields from -100 to 100 mT. $I_c^0$ and $I_c^1$ correspond to the intrinsic and induced superconducting critical current of NbSe$_2$ and $\alpha$-RuCl$_3$, respectively. \textbf{(b)} Current dependent resistance with positive and negative excitations under fields of 0, $\pm$2 and $\pm$5 mT. The lines are plotted near $I_c$ to highlight the nonreciprocal $I_c$. \textbf{(c)} Field dependence $I_c^0$ and $I_c^1$ at both positive and negative branches.  \textbf{(d)} Supercurrent rectification efficiency $Q$=[$I_{c+}$-$I_{c-}$]/[$I_{c+}$+$I_{c-}$] plotted as the function of field $B$. $|Q|$ reaches the maximum value of 15\% at $B$= $\pm$ 3 mT.
}
\end{figure*}

In order to investigate the chiral superconductivity state, we also conducted a study on the nonreciprocal transport properties measurement. As the magnetic field was applied along out-of-plane, a typical second harmonic resistance can be observed as provided in Fig. S9, indicating the existence of inversion symmetry breaking within in-plane.  Therefore, we further investigated the superconducting diode effect (SDE) in the $\alpha$-RuCl$_3$. Fig. \ref{fig4} presents the mapping of d$V$/d$I$ under low magnetic fields, ranging from -100 to 100 mT at 1.8 K. This mapping discloses two transitions, which are associated with the intrinsic critical current of NbSe$_2$ ($I_c^0$) and the superconductivity induced in$\alpha$-RuCl$_3$ ($I_c^1$), as comprehensively explained above. The $I_c^0$ exhibits an even symmetry with respect to both current and field, while $I_c^1$ demonstrates a noticeable asymmetry under a magnetic field ranging from -70 to 70 mT. This suggests that $I_c^1$ behaves nonreciprocally, exhibiting a typical SDE under a finite range of out-of-plane magnetic field and current.

To gain a more comprehensive understanding of the SDE of critical current, we present resistance-current curves under several typical fields of $\pm$ 5,  $\pm$ 2, and 0 mT in Fig. \ref{fig4}(b). Under a positive field of 2 mT, the negative critical current ($I_c^-$) is 750 $\mu$A, whereas the positive critical current ($I_c^+$) is 535 $\mu$A. This demonstrates a nonreciprocal transport in this SDE device, with a nonreciprocity window of approximately 180 $\mu$A. Notably, this asymmetry of the critical current can be reversed by applying a magnetic field of -2 mT, where the values of $I_c^-$ and $I_c^+$ are inverted to 538 and 750 $\mu$A, respectively. The significant difference between the negative and positive current values, expressed as $\Delta I_c$= $I_{c+}$- $\mid I_{c-}\mid$, provides strong evidence of SDE behavior. The sign dependency of $I_c$ exhibits the hallmark behavior of chiral superconductivity with respect to both current and magnetic fields. As the field increases to 5 mT, the SDE is diminished in contrast. Notably, the SDE feature can still be observed even at zero field, suggesting that a weak residual magnetic field may induce the SDE.

To contrast the differences in critical current, the negative and positive critical currents, denoted as $I_c^0$ and $I_c^1$, are graphically represented in Fig. \ref{fig4}(c). Evidently, $I_c^0$ does not exhibit the chiral phenomenon, as it remains independent of both current and field directions. For $I_c^1$, however, there is an asymmetry between the negative $I_{c+}^1$ and the positive $I_{c-}^1$. To further assess the difference between $I_{c+}$ and $I_{c-}$, we introduce an analysis of the superconductivity diode efficiency, defined as $Q$=[$I_{c+}$-$I_{c-}$]/[$I_{c+}$+$I_{c-}$]. The field-dependent $Q$s values for both $I_c^0$ and $I_c^1$ are evaluated, as depicted in Fig. \ref{fig4}(d). While the $Q$ value for NbSe$_2$ is nearly zero, the $Q$ of $\alpha$-RuCl$_3$ is significantly modulated by the magnetic field, particularly in the low field region. At finite weak fields, $Q$ increases almost linearly with the field until it reaches a breakdown field of approximately 3 mT. At this point, the maximum magnitude of $Q$ is observed to be about 14\%, a value comparable to previous reports on Josephson junctions \cite{wu2022field,ghosh2024high,lin2022zero}. We underscore that the diode effect observed in the present study is exceptionally sensitive to the magnetic field, as evidenced by the considerably smaller breakdown field compared to previous results (tens of mT).

When considering the mechanism of SDE, it is crucial to understand the breaking of both spatial-inversion and time-reversal symmetry \cite{nadeem2023superconducting,jiang2022superconducting,mao2024universal,yuan2022supercurrent}. Recent reports of SDE in two-dimensional systems focus on superconducting hybrid structures with magnetic or other quantum materials \cite{ando2020observation,narita2022field}, which are primarily based on the breaking of rotational or time-reversal symmetry on the interface of a two-dimensional superconductor through the interaction with other materials \cite{de2021gate,jeon2022zero}. In two-dimensional limit or interface, the inversion symmetry is naturally broken to be noncentrosymmetric. The existence of strong spin-orbit coupling (SOC) fields will lead to substantial spin splitting between these valleys, where the broken in-plane inversion symmetry could induce polarization of electron spin. Once applying an external magnetic field or with the existence of spontaneously time-reversal symmetry, the nonreciprocal superconductivity will be induced. Among these 2D systems, the Ising- or Rashba-type superconductivity has been widely investigated.

In Ising- or Rashba-type superconductivity, the broken in-plane symmetry can generate an out-of-plane or in-plane spin polarization along with an effective Zeeman field, resulting in a Zeeman-type spin splitting at each pocket \cite{wan2023orbital,yi2022crossover,xi2016ising,wickramaratne2020ising,yoshizawa2021atomic,wang2021ising,de2018tuning}. Given that the spin splitting near the $\Gamma$ point is significantly weaker than that near the $\textbf{K}$ and $\textbf{K'}$ points, the spin-momentum locking effects are predominantly governed by the $\textbf{K}$ and $\textbf{K'}$ pockets \cite{wan2023orbital,yi2022crossover,xi2016ising}. Furthermore, the spin-splitting effects exhibit opposite signs at the $\textbf{K}$ and $\textbf{K'}$ points. Considering the spin polarization directions, once the spin is locked along the $c$-axis, it will lead to unconventional superconductivity pairs as a so-called Ising-type. By applying an out-of-plane external field ($B_z$) onto the Ising system, an additional Zeeman energy, $\Delta_Z$, is introduced, breaking the time-reversal symmetry. The pockets near the $\textbf{K}$ and $\textbf{K'}$ points are further split into 2($\Delta_{SOC}$-$\Delta_Z$) and 2($\Delta_{SOC}$+$\Delta_Z$), respectively. Oppositely, once a broken inversion symmetry occurs along the out-of-plane of a superconductor, mostly in a 2D interface within a heterostructure, SOCs will cause a spin-momentum locking within the in-plane direction, and another unconventional superconductivity can be induced as a so-called Rashba-type \cite{mao2024universal,amundsen2024colloquium,yoshizawa2021atomic}. 

Similar to these systems that combine superconductors and magnetic materials, both the inversion symmetry breaking at the heterojunction interface and the exchange field of $\alpha$-RuCl$_3$ can jointly contribute to the SDE. To explore the vector value of the nonreciprocal supercurrent, it's critical to understand the spatial-inversion and time-reversal symmetry breaking within the $\alpha$-RuCl$_3$ layer. We rotated the magnetic field from the in-plane to the out-of-plane orientation, and found that the SDE only occurs when the field is aligned to the out-of-plane direction. Therefore, the induced superconductivity in $\alpha$-RuCl$_3$ appears to be inconsistent with Rashba-type, but rather aligns with the characteristics of an Ising-type superconductor. Nevertheless, a nature doubt could be aroused that the Ising-like superconductivity is from the NbSe$_2$ rather than from $\alpha$-RuCl$_3$. Therefore, we also investigated the NbSe$_2$/CrCl$_3$ heterostructures as provided in Fig. S14, where few layers of CrCl$_3$ can also induce superconductivity by proximity effect from NbSe$_2$, but without nonreciprocal supercurrent or obvious SDE. While CrCl$_3$ is also an antiferromagnetic insulator, spatial-inversion and time-reversal symmetry breaking can occur for monolayer or few layers of CrCl$_3$, the SOC is considerably weaker than that of $\alpha$-RuCl$_3$ with 3$d$ electronic configuration of Cr$^{3+}$ rather than 4$d$ electrons of Ru$^{3+}$ and nearly octahedral coordination due to quenching of the orbital moment\cite{PhysRevMaterials.1.014001,ran2017prl}. Thus, we can basically conclude that the nonreciprocal supercurrent and SDE are dominated by the Ising-SOC superconductivity in monolayer and few layers $\alpha$-RuCl$_3$ in our present work.

\section*{Conclusion}

To conclude, through the preparation of a heterostructure comprising the magnetic Mott insulator $\alpha$-RuCl$_3$ and $s$-wave superconductor NbSe$_2$, we have detected nonreciprocal superconductivity within $\alpha$-RuCl$_3$, induced by the proximity effect from NbSe$_2$. The induced superconductivity have been found to be considerably weaker than that of pure NbSe$_2$ according to the temperature and field dependencies. Intriguingly, we observed the SDE effect in the heterojunction, with the corresponding rectification ratio being 14$\%$. We anticipate that our findings will provide a novel platform for the development of superconducting electronic devices.

\section*{Methods}

\subsection*{Crystal Synthesis}
Single crystals $\alpha$-RuCl$_3$ were synthesized using the chemical vapor transport method\cite{ran2017prl}. High-purity commercial $\alpha$-RuCl$_3$ powder was sealed in an ampoule in vacuum and subsequently heated in a tube furnace. The growth of single crystals starts with a temperature gradient along the ampoule during a slow cooling process from 700 $^\circ$C. The obtained crystals typically exhibit a plate-like shape. Dispersive x-ray spectroscopy and x-ray Laue diffraction were employed to verify the single crystallinity and lattice direction.

\subsection*{Device Fabrication}
The vdW heterostructure devices were fabricated using dry transfer technique in the atmosphere of argon. Thin pieces of $h$-BN, NbSe$_2$ and few-layer RuCl$_3$ were prepared by exfoliating high-quality single crystals of $h$-BN, NbSe$_2$ and RuCl$_3$, respectively. The NbSe$_2$ and RuCl$_3$ flakes were then picked up with polyvinyl alcohol (PVA) at 75 degrees centigrade, transferred and aligned onto the substrates in sequence, with a $h$-BN flake covering the top to prevent device degradation.

To ensure robust contact between the electrodes and thin flakes, electrodes were pre-fabricated onto SiO$_2$/Si substrates. Firstly, the circuit pattern was written onto the substrates using a Laser Direct-Write lithography system (MicroWriter ML3) within standard lithography process. The electrode pattern was subsequently etched by 25 nm using a Reaction Ion Etching (RIE) system, followed by the deposition of 5 nm of Ti and 25 nm of Au via magnetron sputtering. These embedded electrodes provide a smooth and solid contact, thereby effectively protecting the few-layer $\alpha$-RuCl$_3$ in our devices from air exposure and potential damage by minimizing the drop from electrodes to substrates.

\subsection*{STEM Characterization}
We utilized a double Cs-corrected Scanning Transmission Electron Microscope (STEM, Grand JEM-ARM300F) to analyze the atomic structure. The microscopy was equipped with a cold field-emission gun and operated at an accelerating voltage of 300 kV. To investigate the interfaces from a cross-sectional view, we cut the device using a Focused Ion Beam (FIB, Grand JIB-4700F). The thickness of these thin specimens was approximately 50-100 nm. To study the element distribution, we employed Energy-Dispersive X-ray Spectroscopy (EDS) mapping.

\section*{Data availability}

The data that support the plots within this paper and other findings of this study are available from the corresponding author upon reasonable request.

\section*{Code availability}
The codes that support this study are available from the corresponding author upon reasonable request.

\section*{Acknowledgements}

This research was supported in part by the Ministry of Science and Technology (MOST) of China (No. 2022YFA1603903, 2021YFA1400400), the National Natural Science Foundation of China (Grants No. 12004251, 12104302, 12104303, 12304217, 12225407, 12074174, and 12374162), the Science and Technology Commission of Shanghai Municipality, the Shanghai Sailing Program (Grant No. 21YF1429200), the Science and Technology Commission of the Shanghai Municipality (21JC1405100). Growth of hexagonal boron nitride crystals was supported by the Elemental Strategy Initiative conducted by the MEXT, Japan, Grant Number JPMXP0112101001, JSPS KAKENHI Grant Number JP20H00354 and A3 Foresight by JSPS.




\bibliography{reference.bib}


\end{document}